\documentclass[a4paper,12pt,onecolumn]{article}
\usepackage{amsmath,amsthm,amssymb}
\usepackage{mathtext}
\usepackage[T1,T2A]{fontenc}
\usepackage[utf8]{inputenc}
\usepackage[english]{babel}
\usepackage{graphicx}
\usepackage{amssymb}
\usepackage{amsmath}
\usepackage{setspace}
\usepackage{cite}

\begin{document}

\begin{center}
	{\large
		{\bf 
			Exact solution for one-dimensional spin models with Markov property
		} 
	}
	\vskip0.5\baselineskip{
		\bf 
		Yu. D. Panov
	}
	\vskip0.5\baselineskip{
		Ural Federal University, 620002 Ekaterinburg, 19 Mira street, Russia
	}
	
\end{center}

\begin{abstract}
For one-dimensional spin and pseudospin models that allow mapping to a Markov chain, the free energy of the system at a finite temperature can be expressed in terms of bond concentrations. Minimizing the free energy function makes it possible to obtain an exact solution of a statistical model. A dilute Ising chain with interacting impurities is considered as an example.
\end{abstract}


\section{Introduction}
One-dimensional statistical anisotropic spin models have traditionally attracted much attention due to their conceptual simplicity and solvability. The two- and three-dimensional anisotropic spin models are typically studied using numerical simulation techniques \cite{Murtazaev2022}. Therefore, the exact solutions of one-dimensional models \cite{Mattis1993} are used to test various theoretical concepts and methods in statistical physics, and also provide a basis for understanding the complex behaviour of real physical systems.

For a large number of anisotropic spin systems, including decorated Ising chains and one-dimensional Blume-Capel, Blume-Emery-Griffiths, and Potts models, the standard approach to obtaining an exact solution involves constructing a transfer matrix and finding its maximum eigenvalue.

The existence of a finite-size transfer matrix allows us to map the model onto a certain Markov chain. This, in turn, allows us to describe in detail various properties of the ground state phases  \cite{Panov2020,Panov2022,PanovRojas2023,Panov2023,Panov2024,Yasinskaya2024}. However, it also allows us to express the free energy of a system at a finite temperature in terms of pair distribution functions or bond concentrations.
 
In this paper, we propose a method for finding an exact solution of a statistical model by minimizing free energy with respect to bond concentrations. As an example, we use a simple one-dimensional model with a frustrated ground state, namely a dilute Ising chain with interacting impurities.

\section{Mapping a one-dimensional spin model onto a Markov chain}

For a certain class of spin and pseudospin chain models, the eigenstates of the system are the direct product of the states at the sites of the chain. In this case, the calculation of the partition function can be done using the Kramers-Wannier transfer matrix, denoted by $\mathbf{W}$. The matrix $\mathbf{W}$ is non-negative, and the free energy of the system, denoted by $f$, in the thermodynamic limit can be expressed in terms of its largest eigenvalue, $\lambda_1$.
\begin{equation}
	f = - \frac{1}{\beta} \ln \left( \lambda_{1} \right) .
\end{equation}
Here, $\beta = 1/(k_{B} T)$, $k_{B}$ is the Boltzmann constant, $T$ is the temperature. 

Each non-negative matrix $\mathbf{W}$ can be mapped to a transition matrix $\mathbf{P}$ of some Markov chain \cite{Gantmakher2000} with matrix elements
\begin{equation}
	P_{ab} = \frac{W_{ab} v_{b}}{\lambda_1 \, v_{a}} ,
\end{equation}
where $W_{ab}$ are matrix elements of $\mathbf{W}$, $v_{a}$ are components of the eigenvector $\left|\lambda_1\right\rangle$, and for a positive matrix, the values of $v_{a}$ can be chosen positive by Perron's theorem~\cite{Gantmakher2000}. Thus, it is possible to map a one-dimensional spin model onto a Markov chain. This fact is well known in the theory of Gibbs measures \cite{Georgii2011}. The opportunity to map a model onto a Markov chain will be called the Markov property.

It is worth to note that for a given model, the matrix $\mathbf{W}$ can be defined in different ways and even have different dimensions. Generally speaking, this will correspond to mappings to various Markov chains. If we define the matrix elements of the transfer matrix as
\begin{equation}
	W_{ab} = e^{ -\beta\left( \frac{\epsilon_{a} + \epsilon_{b} }{2} + \epsilon_{ab} \right) } ,
\end{equation}
where $\epsilon_{a}$ is the energy of the $a$ site state, $\epsilon_{ab}$ is the energy of interaction of neighboring sites in states $a$ and $b$.
This choice of $\mathbf{W}$ is convenient because the limiting state of the Markov chain directly gives concentrations $n_{a}$ of the number of chain sites in the state $a$, $a=1,\ldots q$. 
With a different choice of $\mathbf{W}$, which may be associated with partial summation over states, it can be difficult to interpret the limiting state of the Markov chain.

An important consequence of mapping the spin model to a Markov chain is the expression of the entropy of the system $\mathcal{S}$ in terms of bond concentrations $x_{ab}$. Let's assume that in the equilibrium state of a spin chain of $N$ sites with periodic boundary conditions there are $N_{ab}$ pairs of nearest neighboring sites in state $a$ on the left site and state $b$ on the right site. We define the concentrations of bonds using the following expressions:
\begin{equation}
	x_{aa} = \frac{ N_{aa} }{N} , \quad
	x_{ab} = \frac{ N_{ab}+N_{ba} }{N} , \; a < b ,
	\label{eq:X}
\end{equation}
then 
\begin{equation}
	\sum_{a, b; a \leq b} x_{ab} = 1 . 
	\label{eq:sumX}
\end{equation}
In total, there are $q(q+1)/2$ concentrations $x_{ab}$, $a\leq b$, where $q$ is the number of site states. Obviously, $x_{ab}$ can be expressed in terms of the pair distribution functions $\left\langle \hat{\delta}_{a,i} \hat{\delta}_{b,i+1} \right\rangle$, where $\hat{\delta}_{a,i}$ is the projector to the $a$ state at the site $i$, $\left\langle \ldots \right\rangle$ means the thermodynamic average:
\begin{equation}
	x_{aa} = \left\langle \hat{\delta}_{a,i} \hat{\delta}_{a,i+1} \right\rangle , \quad
	x_{ab} = \frac{1}{2} 
	\left( \left\langle \hat{\delta}_{a,i} \hat{\delta}_{b,i+1} \right\rangle 
	+ \left\langle \hat{\delta}_{b,i} \hat{\delta}_{a,i+1} \right\rangle\right) , \; a < b .
	\label{eq:X-PDF}
\end{equation}

The Markov property leads to the following expression for the entropy of the spin chain \cite{Panov2022}:
\begin{equation}
	\mathcal{S} = - \sum_{a, b; a \leq b} x_{ab} \ln x_{ab} 
	+ \ln2 \sum_{a, b; a < b} x_{ab}
	+ \sum_{a} n_{a} \ln n_{a} , 
	\label{eq:S}
\end{equation}
where 
\begin{equation}
	n_{a} = x_{aa} 
	+ \frac{1}{2} \sum_{b; a>b} x_{ba} 
	+ \frac{1}{2} \sum_{b; a<b} x_{ab} .
\end{equation}
The expression \eqref{eq:S} allows us to calculate the residual entropy of a spin chain with the Markov property and find the states of the system at the boundaries of the ground state phases in the phase diagram  \cite{Panov2022,Panov2023,Panov2024,Yasinskaya2024}.

For the class of spin models under consideration, the internal energy of the system can also be expressed in terms of bond concentrations:
\begin{equation}
	\varepsilon = \frac{1}{N} \left\langle \hat{H} \right\rangle = \varepsilon(x_{ab}) , 
\end{equation}
where $\hat{H}$ is the Hamiltonian of the system. As a result, for a spin model with the Markov property, it is possible to find all the thermodynamic characteristics without calculating the largest eigenvalue of the transfer matrix, but minimizing the free energy of the system
\begin{equation}
	f(x_{ab}) = - \varepsilon(x_{ab}) + T \mathcal{S}(x_{ab})
\end{equation}
with respect to the variables $x_{ab}$, taking into account the corresponding constraints, in particular \eqref{eq:sumX}. 

Next, we will demonstrate this possibility using the example of a dilute Ising chain with interacting nonmagnetic impurities. 

\section*{Dilute Ising chain}

The Hamiltonian of a one-dimensional dilute Ising chain with interacting nonmagnetic impurities can be written as follows:
\begin{equation}
	\hat{H} = -J \sum_{i} \hat{S}_{z,i} \hat{S}_{z,i+1} 
	+ V \sum_{i} \hat{\delta}_{0,i} \hat{\delta}_{0,i+1} 
	- h \sum_{i} \hat{S}_{z,i} .
	\label{eq:Ham}
\end{equation}
We use the pseudospin $S=1$ operator, where the states of the spin doublet and the nonmagnetic impurity correspond to the $z$ projections $S_z = \pm1$ and $S_z = 0$, $J$ is the exchange integral, $V>0$ is an effective interstitial interaction for impurities, $h$ is an external magnetic field. We use an annealed version of the model when positions of impurities are not fixed in the chain. Model \eqref{eq:Ham} in a zero magnetic field has an exact solution~\cite{Rys1969}, which is most generally analyzed in the work~\cite{Balagurov1974}. The properties of local distributions of states at the nodes of the chain were studied in reference~\cite{Panov2020}, and the entropy and phase diagram of the ground state of the system in a magnetic field with a fixed concentration of impurities were obtained in ~\cite{Panov2022}.

Let's introduce the concentrations of bonds $x_{1,1}$, $x_{-1,-1}$, $x_{0,0}$, $x_{0,1}$, $x_{0,-1}$, $x_{1,-1}$ in accordance with the expressions \eqref{eq:X}. For the internal energy of the chain per site, we get
\begin{equation}
	\varepsilon = -J \left( x_{1,1} + x_{-1,-1} - x_{1,-1} \right) + V x_{0,0} 
	- h \left[ x_{1,1} - x_{-1,-1} + \frac{1}{2} \left(  x_{0,1} - x_{0,-1} \right) \right] . 
\end{equation}
Along with the restriction \eqref{eq:sumX}, we will assume that the concentration of impurities $n$ is fixed:
\begin{equation}
	n = x_{00} + \frac{1}{2} \left( x_{0,1} + x_{0,-1} \right) = const .
	\label{eq:n}
\end{equation}

As a result, the problem is reduced to finding the maximum of the function
\begin{equation}
	\varphi = - \beta \varepsilon + \mathcal{S} 
	+ \lambda \left( \sum_{a, b} x_{ab} - 1\right) 
	+ \mu \left(  x_{00} + \frac{1}{2} \left( x_{0,1} + x_{0,-1} \right) - n\right) . 
\end{equation}
Here $\lambda$ and $\mu$ are Lagrange multipliers, $\mathcal{S}$ is defined by the equation \eqref{eq:S}.

Using the necessary extremum conditions for $\varphi$, we obtain the following equations:
\begin{equation}
	\beta \left( J \pm h \right) - \ln x_{\pm 1,\pm 1} + \ln n_{\pm 1} + \lambda = 0 , 
	\label{eq:eqX11}
\end{equation}
\begin{equation}
	- \beta J - \ln x_{1,-1} + + \frac{1}{2}\ln n_{1} + \frac{1}{2}\ln n_{-1} + \ln2 + \lambda = 0 , 
	\label{eq:eqX1-1}
\end{equation}
\begin{equation}
	-\beta V - \ln x_{0,0} - 1 + \lambda + \mu = 0 , 
	\label{eq:eqX00}
\end{equation}
\begin{equation}
	\pm \frac{1}{2} \beta h - \ln x_{0,\pm 1} + \frac{1}{2}\ln n_{\pm 1} 
	- \frac{1}{2} + \ln2 + \lambda + \frac{\mu}{2} = 0 , 
	\label{eq:eqX01}
\end{equation}
and the constraints \eqref{eq:sumX} and \eqref{eq:n}. Here, $n_{\pm1} = x_{\pm 1,\pm 1} + (x_{1,-1} + x_{0,\pm 1})/2$.

Transformations of this system of equations lead to fourth-degree equations for individual $x_{ab}$, the solutions of which, in analytical form, are very complex. A similar issue arises when using the conventional approach, which involves finding the largest eigenvalue of the transfer matrix. However, even in the absence of an explicit solution, it is possible to obtain from the equations (\ref{eq:eqX11}--\ref{eq:eqX01}) a number of relations between the quantities $x_{ab}$ and $n_{a}$ that have a clear physical meaning:
\begin{equation}
	\frac{x_{1,-1}^{2}}{x_{1,1} x_{-1,-1}} = 4 e^{-4\beta J} , \quad
	\frac{x_{1,1}}{x_{-1,-1}} = \frac{x_{0,1}^{2}}{x_{0,-1}^{2}} , \quad
	\frac{x_{1,1}n_{-1}}{x_{-1,-1}n_{1}} = e^{-2\beta h} .
\end{equation}
Next, we will consider analytical solutions for two special cases, namely zero impurity concentration and zero magnetic field.

\subsection{Simple Ising chain, $n=0$}

In the absence of impurities, $n=0$, we have $x_{0,0}=0$, $x_{0,1}=0$, $x_{0,-1}=0$, and the equations \eqref{eq:eqX00} and \eqref{eq:eqX01} disappear. The combination of the equations \eqref{eq:eqX11} and \eqref{eq:eqX1-1} leads to the relations:
\begin{equation}
	\frac{x_{1,-1}^{2}}{x_{1,1} x_{-1,-1}} = \rho, \quad 
	\rho = 4 e^{-4\beta J} , 
\end{equation}
\begin{equation}
	\frac{x_{-1,-1} n_{1}}{x_{1,1} n_{-1}} = \gamma, \quad 
	\gamma = e^{-2\beta h} . 
\end{equation}
Taking into account \eqref{eq:sumX}, we get
\begin{equation}
	x_{1,1} = \frac{\phi - \gamma + 1}{\phi \left( \phi + \gamma + 1\right) } , 
\end{equation}
\begin{equation}
	x_{-1,-1} = \frac{\gamma \left( \phi + \gamma - 1 \right) }{\phi \left( \phi + \gamma + 1\right) } ,
\end{equation}
\begin{equation}
	x_{1,-1} = \frac{ \gamma \rho }{\phi \left( \phi + \gamma + 1\right) } ,  
\end{equation}
where 
\begin{equation}
	\phi = \sqrt{\left( \gamma - 1 \right)^{2} + \gamma \rho } . 
\end{equation}

The solutions obtained lead to well-known expressions for the thermodynamic functions of the Ising chain.
For example, for the magnetization we have:
\begin{equation}
	M = x_{1,1} - x_{-1,-1} 
	= \frac{\sinh( \beta h ) }{\sqrt{\sinh^{2}( \beta h ) + e^{-4\beta J}}} . 
\end{equation}

\subsection{The case of a zero magnetic field}

When $h=0$, the number of equations is also reduced, because $x_{1,1}=x_{-1,-1}$, $x_{0,1}=x_{0,-1}$, and $n_{1}=n_{-1}$. In this case, it is convenient to enter the deviation of the impurity concentration from the half filling, $m = n - 1/2$, and then
\begin{equation}
	x_{0,0} + x_{0,1} = \frac{1}{2} + m  . 
\end{equation} 
The combinations of the equation (\ref{eq:eqX11}-\ref{eq:eqX1-1}) allow us to write the relations
\begin{equation}
	x_{1,-1} = 2 e^{-\beta J} x_{1,1} , 
\end{equation}
\begin{equation}
	x_{0,1}^2 = 4 e^{\beta(V-J)} x_{0,0} x_{1,1} . 
\end{equation}
Taking into account the equation \eqref{eq:sumX}, we also obtain
\begin{equation}
	2 x_{1,1} + x_{1,-1} + x_{0,1} = \frac{1}{2} - m  . 
\end{equation}

Solving these equations, we find:
\begin{equation}
	x_{1,1} = \frac{ e^{\beta J} \left( 1-2m \right) \left( g-m\right) 
	}{ 4\cosh(\beta J)\left( 2g+1\right)  } ,
\end{equation}
\begin{equation}
	x_{0,0} = \frac{  \left( 1+2m \right) \left( g+m\right) }{ 2g+1 } ,
\end{equation}
\begin{equation}
	x_{1,-1} = \frac{ e^{-\beta J} \left( 1-2m \right) \left( g-m\right) 
	}{ 2\cosh(\beta J)\left( 2g+1\right)  } ,
\end{equation}
\begin{equation}
	x_{0,1} = \frac{   1-4m^{2}  }{ 2\left( 2g+1 \right) } ,
\end{equation}
where 
\begin{equation}
	g = \sqrt{m^{2} + \left( \frac{1}{4}-m^{2} \right) e^{-\beta V}\cosh(\beta J)} . 
\end{equation}
Taking into account the relationships between the concentrations of bonds and the pair distribution functions \eqref{eq:X-PDF}, these expressions exactly match those found in reference \cite{Panov2020}.

\section{Сonclusion}
The paper considers the features of the thermodynamic description of spin and pseudospin chains, for which the partition function can be expressed in terms of the Kramers-Wannier transfer matrix. 
The standard method for calculating the free energy in this situation involves finding the largest eigenvalue of the transfer matrix. 
Another possibility is to use the mapping of models of this class to a Markov chain. In this case, the free energy can be expressed as a function of bond concentrations. 
The paper presents the general structure of this function. 
Using the example of a dilute Ising chain, we show that solving the problem of finding the conditional extremum of this function provides an exact solution for the statistical model. 
This allows us to determine all the thermodynamic properties of the system.
The considered method can be used for other anisotropic spin systems with Markov property, including decorated Ising chains and one-dimensional Blume-Capel, Blume-Emery-Griffiths, and Potts models.

\subsection*{Funding}
The work was supported by the Ministry of Science and Higher Education of the Russian Federation, the project FEUZ-2023-0017.


\end{document}